\documentclass{ws-p8-50x6-00}
\usepackage{epsfig}

\renewcommand{\today}{1.~September~2000}

\begin{document}

\title{Proton Emission Times in Spectator Fragmentation}

\author{C. Schwarz for the ALADIN collaboration}

\address{Gesellschaft  f\"ur  Schwerionenforschung, D-64291 Darmstadt, Germany}

\maketitle

\abstracts{
Proton-proton correlations from spectator decays following
$ ^{197}$Au+$^{197}$Au collisions at 1000 AMeV have been measured with
an highly efficient detector hodoscope. The constructed correlation
functions indicate a moderate expansion and low breakup densities
similar to assumptions made in statistical multifragmentation models.
In agreement with a volume breakup rather short time scales were deduced
employing directional cuts in proton-proton correlations.
}

\section{Introduction}
Densities lower than ground state
density of nuclei are a prerequisite of statistical models describing 
multifragmentation \cite{Bondorf951,Gross972}. 
While static statistical models assume fragment formation in an expanded volume 
breakup, the dynamic statistical model\cite{Friedman837,Friedman901} 
combines surface emission during expansion with volume breakup of the remaining source.
Interferometry-type methods are widely considered
as valuable tools in determining the space-time extent of such sources and, recently,
spectator remnants of the reaction Au + Au at 1 AGeV incident energy were found
to break up at densities considerably lower than ground state density \cite{Fritz995}.
In that analysis a instantaneous volume breakup was assumed.

In this article we report on results of a directional analysis of the measured 
proton-proton correlations of protons from the target spectator following
the collisions of Au + Au at 1 AGeV incident energy. 
The results are found to be consistent
with low breakup densities with values close to those assumed in the statistical
multifragmentation models and short emission time differences with values
close to those anticipated for volume breakup.

\section{Experiment}

Targets, consisting of $25$ mg/cm$^{2}$ of $^{197}$Au were irradiated
by an 1 AGeV Au beam delivered by the Schwerionen-Synchrotron (SIS) at GSI in
Darmstadt. 
For the results presented here, we employed one multi-detector
hodoscope, consisting of a total of 96 Si-CsI(Tl) telescopes in closely packed
geometry. The hodoscope covered an angular range $ \Theta _{lab} $ from 122$ ^{0} $
to 156$ ^{0} $ with the aim of selectively detecting the products of the
target-spectator decay. Each telescope consisted of a 300 \mbox{$\mu$m}  Si detector
with \mbox{30 x 30 mm$^{2}$} active area, followed by a 6 cm long CsI(Tl) scintillator
with photodiode readout. The distance to the target was 60 cm. The products
of the projectile decay were measured with the time-of-flight wall of the ALADIN
spectrometer \cite{Schuettauf961} and the quantity $ Z_{BOUND} $ was determined
event-by-event. $ Z_{BOUND} $ is defined as the sum of the atomic numbers
$ Z_{i} $ of all fragments with $ Z $$ _{i} $$ \geq 2 $. It reflects
the variation of the charge of the primary spectator system and serves as a
measure of the impact parameter.

\section{Data analysis}

The correlation functions were constructed dividing the spectrum of relative
momenta of two coincident particles by the spectrum of pairs from different events. 
At a relative momentum of $ q\approx 20 $ MeV/c, one observes a maximum of the 
correlation function the height of which is inversely related to the diameter 
of the source for simultaneous emission \cite{Koonin779,Pratt871}.
Directional cuts on the angle between 
sum momentum and relative momentum allow the determination of the spatial and temporal
separation of the two protons at emission\cite{Koonin779,Lisa933}.
Instead of using sharp cuts on the angle between sum momentum and relative momentum we 
employed harmonic weights ($sin^2$ and $cos^2$) for the generation of transversal
and longitudinal correlation functions\cite{Schwarz001}. 
%---------------------------------------------------------------------------------
\begin{figure}[tbh]
\begin{centering}
\includegraphics*[angle=-90,scale=0.6]{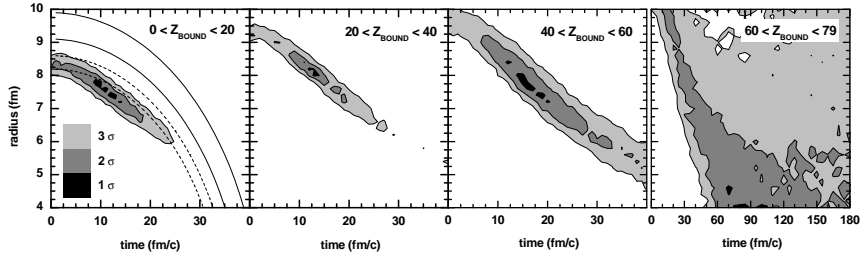}
\caption{\small \label{maps}
$\chi^2$ distributions as a function of radius and emission time of the 
emitting source for four impact parameter ranges indicated by the $Z_{BOUND}$ ranges
in the panels. The lines in the left panel are explained in the text.}
\end{centering}
\end{figure}
%---------------------------------------------------------------------------------
The analysis of the p-p correlation functions was performed with the Koonin-Pratt 
formalism \cite{Koonin779,Pratt871}. Particles were chosen to be randomly emitted 
from the volume of an uniform sphere and their velocities were sampled according 
to a Maxwell-Boltzmann distribution. 
An additional velocity component was added in order to simulate the 
Coulomb repulsion corresponding to the location of the particle within the source.
An experimentally observed anisotropy in the proton-energy spectra was modeled
by assuming a sideward movement (bounce) of the source perpendicular to the beam.
This causes a reaction plane which was included in the Monte-Carlo calculations.

\section{Results and Discussions}

We varied the radius of an uniform density distribution and the Gaussian emission 
time of the protons. The simulated correlation functions were used to perform 
a $\chi^2$-test in the region of relative momentum region  
\mbox{$10 \le q \le 35$ MeV/c}. The results are presented in Fig. \ref{maps}, where the 
shadings denote the deviation from the data. 
The minima of $\chi^2$ yield approximately constant source radii of 
\mbox{$\approx 8$ fm} and short emission times of \mbox{$\approx 10-15$ fm/c}. 
For the most peripheral bin a radius and emission 
time could not be deduced due to the low statistics of the correlation function. 
The experimental correlation functions (symbols) and the simulations (lines) for the 
parameters corresponding to the minimum of $\chi^2$ are shown 
%------------------------------------------------------------------------------------
\begin{figure}[bth]
\begin{minipage}[h]{8cm}
\includegraphics*[scale=0.28,angle=-90]{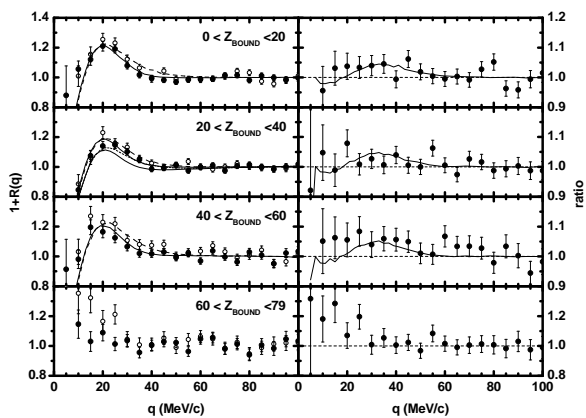}
\end{minipage}
\begin{minipage}[b]{3.5cm}
\parbox{3.4cm}
{
%\vspace{-1cm}
\caption{\small \label{datasim}
The left panel shows experimental longitudinal (open symbols) and 
transversal (closed symbols) correlation functions and results of MC-simulations (lines). 
The right panel compares the their experimental ratios (symbols) with the results of the 
MC-simulation (lines).}
}
\end{minipage}
\end{figure}
%------------------------------------------------------------------------------------
in Fig. \ref{datasim}, left panel.  
They agree well with each other. The minimum \mbox{$\chi^2$} values per degree of 
freedom are within the range of \mbox{$1.3 \le \chi^2/d.o.f. \le 2.4$}. 
The right panel in Fig. \ref{datasim} presents the ratios between longitudinal and 
transversal correlation functions. One recognizes the weak enhancement of about 5\% 
above unity (dashed line) of data (symbols) and simulations (solid line) 
for relative momenta $q<50$ MeV/c. For a source size of \mbox{$\approx 8$ fm} the quantum
suppression is expected at 
\mbox{$q=\sqrt{{\bf q}^2}<\sqrt{3}*\hbar/r \approx 40$ MeV/c}.
  
\section{Conclusions}

We constructed correlation functions for pairs of protons detected at backward angles
in the reaction Au+Au at 1000 AMeV incident beam energy. Using high energy cuts of
\mbox{$E>20$ MeV} we selected protons which are only little affected by sequential 
feeding \cite{Schwarz991}.
Comparing the results with Monte-Carlo simulations within the Koonin-Pratt formalism 
fairly constant freeze-out radii of $R\approx8$ fm are deduced and 
emission times of $\tau=10-15$ fm/c are surprisingly short of the order of the passing 
time of the projectile through the target.
The extracted radii are larger than the ground state
radii of target spectators and show expansion. 
Because of the short emission times in the order
of the passing time of both spectators we cannot exclude that the protons
come from first stage scattering of the nuclear cascade.

\section*{Acknowledgments}
This work was supported by the Deutsche Forschungsgemeinschaft 
under contract Schw510/2-1 and 
by the European Community under contract ERBFMGECT-950083.

\end{document}